\documentclass[aps,prl,twocolumn,amsmath,amssymb,nofootinbib,superscriptaddress,floatfix,reprint,longbibliography]{revtex4-1}
\usepackage[dvips]{graphicx}
\usepackage{latexsym}
\usepackage{amsmath}
\usepackage{amsfonts}
\usepackage{amssymb}
\usepackage{color}
\usepackage{txfonts}
\usepackage{float}
\usepackage{url}
\usepackage[colorlinks=true, urlcolor=blue, linkcolor=blue, citecolor=blue, pdftex]{hyperref}
\usepackage{ulem}
\usepackage{graphicx}
\usepackage{extpfeil}
\usepackage{subfigure}
\usepackage{physics}
\DeclareUnicodeCharacter{2212}{-}

\begin{document}
	\newcommand{\fig}[2]{\includegraphics[width=#1]{#2}}
	\newcommand{\la}{{\langle}}
	\newcommand{\ra}{{\rangle}}
	\newcommand{\dg}{{\dagger}}
	\newcommand{\upa}{{\uparrow}}
	\newcommand{\dna}{{\downarrow}}
	\newcommand{\ab}{{\alpha\beta}}
	\newcommand{\ias}{{i\alpha\sigma}}
	\newcommand{\ibs}{{i\beta\sigma}}
	\newcommand{\hH}{\hat{H}}
	\newcommand{\hn}{\hat{n}}
	\newcommand{\hc}{{\hat{\chi}}}
	\newcommand{\hU}{{\hat{U}}}
	\newcommand{\hV}{{\hat{V}}}
	\newcommand{\br}{{\bf r}}
	\newcommand{\bk}{{{\bf k}}}
	\newcommand{\bq}{{{\bf q}}}
	\def\gsim{~\rlap{$>$}{\lower 1.0ex\hbox{$\sim$}}}
	\setlength{\unitlength}{1mm}
	\newcommand{{\vhf}}{$\chi^\text{v}_f$}
	\newcommand{{\vhd}}{$\chi^\text{v}_d$}
	\newcommand{{\vpd}}{$\Delta^\text{v}_d$}
	\newcommand{{\ved}}{$\epsilon^\text{v}_d$}
	\newcommand{{\vved}}{$\varepsilon^\text{v}_d$}
	\newcommand{\pprl}{Phys. Rev. Lett. \ }
	\newcommand{\pprb}{Phys. Rev. {B}}

\title {The electronic and magnetic structures of bilayer La$_3$Ni$_2$O$_7$ at ambient pressure}
\author{Yuxin Wang}
\affiliation{Beijing National Laboratory for Condensed Matter Physics and Institute of Physics,
	Chinese Academy of Sciences, Beijing 100190, China}
\affiliation{School of Physical Sciences, University of Chinese Academy of Sciences, Beijing 100190, China}

\author{Kun Jiang}
\email{jiangkun@iphy.ac.cn}
\affiliation{Beijing National Laboratory for Condensed Matter Physics and Institute of Physics,
	Chinese Academy of Sciences, Beijing 100190, China}
\affiliation{School of Physical Sciences, University of Chinese Academy of Sciences, Beijing 100190, China}

\author{Ziqiang Wang}
\email{wangzi@bc.edu}
\affiliation{Department of Physics, Boston College, Chestnut Hill, MA 02467, USA}

\author{Fu-Chun Zhang}
\affiliation{Kavli Institute of Theoretical Sciences, University of Chinese Academy of Sciences,
	Beijing, 100190, China}
\affiliation{Collaborative Innovation Center of Advanced Microstructures, Nanjing University, Nanjing 210093, China}

\author{Jiangping Hu}
\email{jphu@iphy.ac.cn}
\affiliation{Beijing National Laboratory for Condensed Matter Physics and Institute of Physics,
	Chinese Academy of Sciences, Beijing 100190, China}
\affiliation{Kavli Institute of Theoretical Sciences, University of Chinese Academy of Sciences,
	Beijing, 100190, China}
 \affiliation{New Cornerstone Science Laboratory, 
	Beijing, 100190, China}

\date{\today}

\begin{abstract}
We carry out a systematic study of the electronic and magnetic structure of the ambient-pressure bilayer La$_3$Ni$_2$O$_7$.
Employing the hybrid exchange-correlation functional, we show that the exchange-correlation pushes the bonding $d_{z^2}$ bands below the Fermi level to be fully occupied. The calculated Fermi surfaces and the correlation normalized band structure match well with the experimental findings at ambient pressure. Moreover, the electronic susceptibility calculated for this new band structure features nesting-induced peaks near the wave vector $Q=(\pi/2, \pi/2)$, suggesting a possible density wave instability in agreement with recent experiments. Through a mean field study and DFT+U calculation introducing a Hubbard U interaction within conventional density functional theory, we confirm the spin-charge intertwined double stripe order is the magnetic ground state. Our results provide a faithful description for the low-pressure La$_3$Ni$_2$O$_7$ electronic structure. 
\end{abstract}
\maketitle

\section{INTRODUCTION}
Being a counterpart to copper, there has long been a theoretical proposition that nickelates could harbor high-temperature superconductivity (SC) \cite{rice_PhysRevB.59.7901,pickett_PhysRevB.70.165109,Khaliullin_PhysRevLett.100.016404,hujp_PhysRevX.5.041012,jphugene2016}, mirroring the physics observed in the high-$T_c$ cuprates \cite{doping_mott,keimer_review,sc_wtcu, iron1,iron2,iron_review}. However, identifying a suitable nickelate proves to be a complex challenge. 
Owing to sustained synthesis efforts \cite{nickelates0,nickelates1,nickelates2,nickelates4,nickelates5}, the manifestation of superconductivity has been successfully accomplished in thin films of the ``infinite-layer" nickelates (Sr,Nd)NiO$_2$ in 2019 \cite{lidanfeng,lidanfeng2,lidanfeng3}, opening the Nickel age of superconductivity \cite{norman}.

In 2023, single crystals of a new type of bulk transition metal oxide La$_3$Ni$_2$O$_7$ (LNO), structurally closer to hole-doped cuprates, were successfully synthesized \cite{meng_wang,meng_wang2,chengjg_crystal}. Under high pressure, a high-temperature superconducting transition $T_c\sim80$K was reported \cite{meng_wang,chengjg,yuanhq,chengjg_crystal,chengjg_poly,sunll}, adding a new family to the nickelates superconductors.
Following its discovery, extensive theoretical studies have been dedicated to understanding this novel material \cite{yaodx,dagotto1,wangqh,Kuroki,guyh,zhanggm,werner,yangf,wucj,dagotto2,yangyf2,ryee2023critical,kun_cpl,PhysRevB.108.L201121,PhysRevB.109.115114}. 
However, the electronic structure of LNO and its SC origin are still under debate. 
Recently, a spin density wave order near $Q=(\pi/2,\pi/2)$ at ambient pressure is confirmed 
from resonant inelastic X-ray scattering (RIXS) \cite{rixs}, muon spin rotation ($\mu$SR) \cite{musr,musr2}, and nuclear magnetic resonance (NMR) \cite{nmr} experiments below 153K \cite{meng_wang2}.
Hence, the low-pressure electronic structure and magnetic structures are highly unexplored. 


In this work, we carry out a systematic study of the electronic and magnetic structure of LNO.  
We find that the crystal field splitting between the two $e_g$ orbitals and the bilayer coupling were highly underestimated in the previous density functional theory (DFT) calculations.  The important exchange-correlation effects turn out to be much stronger, resulting in large splittings that push the bonding $d_{z^2}$ bands below the Fermi level to become fully occupied. This leads to distinct new electronic structures for LNO. The calculated Fermi surfaces and the correlation normalized band structure match well with ARPES measurements at ambient pressure. In contrast, previous DFT results have not described this very well. Based on our new results, we find that LNO at ambient pressure harbors the density wave order close to the wave vector $Q$. Then, using the mean-field decoupling and DFT+U methods, we successfully identify the double stripe magnetic structure of LNO at ambient pressure. 

\begin{figure}
	\begin{center}
		\fig{3.4in}{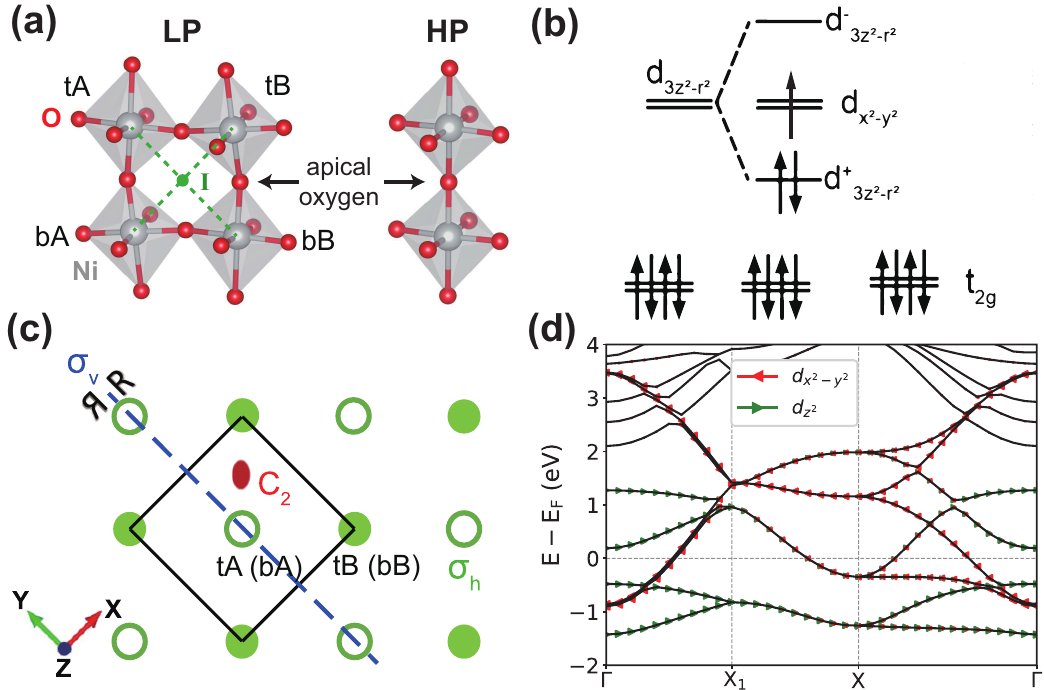}
		\caption{(a) The essential part of the La$_3$Ni$_2$O$_7$ atomic structure is the bilayer NiO$_2$ plane formed by two corner-shared NiO$_6$ octahedrons. For the low-pressure (LP) phase, the octahedrons are tilted alternatively resulting in 4 Ni atoms in each unit cell. There are four Ni atoms per unit cell, which are labeled as tA, tB, bA, bB with t/b as layer index and A/B as sublattice index. The inversion center is indicated by I. For the high-pressure (HP) phase, the bilayer octahedrons line up with 2 Ni atoms in each unit cell. The apical oxygens are crucial for the electronic structure. (b) The crystal field and the occupation configuration in the (2Ni)$^{+5}$ state.
        The $t_{2g}$ orbitals are fully occupied. The strong bilayer coupling mediated by the apical oxygen strongly splits the $d_{z^2}$ energy level into bonding and antibonding orbitals. 
        (c) The symmetry generators in the LP phase of LNO. (d) The band structure of La$_3$Ni$_2$O$_7$ in the LP phase calculated using the HSE06 exchange-correlation functional.}
    \label{fig1}
	\end{center}
\end{figure}

\section{CRYSTAL STRUCTURE}
Much like the cuprates, the essential part of the LNO superconductor is the bilayer NiO$_2$ block \cite{meng_wang}, as illustrated in Fig. \ref{fig1}(a).  We label them as the top (t) and bottom (b) layers. Around each Ni site, six oxygen atoms form a standard NiO$_6$ octahedron. The two nearest neighbor octahedrons between the two layers are corner-shared by one apical oxygen. The LNO at ambient pressure is in the $Amam$ phase with the two octahedrons tilted in an alternating fashion, leading to four Ni atoms inside each unit cell. We label them as tA, tB, bA, bB with t/b indicating the layer index and A/B the sublattice index. This low-pressure (LP) phase evolves into the high symmetry $Fmmm$ structure phase under high pressure (HP). The two octahedrons line up with two Ni atoms per unit cell as shown in Fig. \ref{fig1}(a) and high-$T_c$ superconductivity emerges. 

For both the $Amam$ and $Fmmm$ structures, the octahedron crystal field splits the Ni $3d$ energy level into the standard $e_{g}$ and $t_{2g}$ complexes, as illustrated in Fig. \ref{fig1}(b). 
Counting the chemical valence in LNO, Ni is in the (2Ni)$^{5+}$ state (Ni$^{2.5+}$ per-site). Hence, the (2Ni)$^{5+}$ has fully occupied $t_{2g}$ orbitals and the $e_g$ orbitals host three electrons.
Furthermore, the apical oxygen plays a crucial role for the electronic structure in LNO.
The strong bilayer coupling between the $d_{z^2}$ orbitals in the top and bottom layers is mediated by the apical oxygen and strongly splits the $d_{z^2}$ energy level into bonding and antibonding orbitals while the $d_{x^2-y^2}$ remains nearly degenerate, as illustrated in Fig. \ref{fig1}(b). 

 The inversion symmetry center $I$ is at the center of four Ni atoms, as shown in Fig. \ref{fig1}(a). For the $Amam$ space group, there are three important symmetry generators, $C_2$ rotation, mirror plane $\sigma_h$, and the mirror plane $\sigma_v$, as illustrated in Fig. \ref{fig1}(c). The $C_2$ operation is along the z direction and exchanges $A$ and $B$ sites. The $\sigma_h$ lies between the top and bottom layers, which maps the $t$ and $b$. Combining $C_2$ and $\sigma_h$ generates the inversion operation $I$. These three operations form the C$_{2h}$ point group of the $Amam$ space group. The other four important symmetry operations in the $Amam$ space group can be generated through the $\sigma_v$ along the $A-A$ direction.

\section{DFT RESULTS}
The first-principles DFT calculation and its extensions are standard ways for determining the electronic structures and the crystal field energies. Various DFT calculations have been applied to finding the electronic structure of LNO \cite{meng_wang,yaodx,guyh,dagotto1}, even long before the discovery of superconductivity in the HP phase \cite{Pickett}. These calculations show that the bonding $d_{z^2}$ bands are active and cross the Fermi level ($E_F$) while the antibonding $d_{z^2}$ bands are empty, indicating an important role of $d_{z^2}$ orbitals. 
On the contrary, recent ARPES measurements find the bonding $d_{z^2}$ bands are below $E_F$ \cite{zhouxj}, which deviates from previous calculations.
We notice that most of the above-mentioned DFT calculations use the Generalized Gradient Approximation (GGA) exchange-correlation functional or its Perdew-Burke-Ernzerhof (PBE) version \cite{perdew1996generalized}.
However, it is widely known that the suitable exchange-correlation functional $E_{xc}$ is crucial for determining band gaps and binding energies \cite{martin2020electronic}. For example, the GGA functional systematically underestimates band gaps of semiconductors \cite{martin2020electronic,exchange_correlation}. 
Because an accurate determination of the positions of the $d_{z^2}$ bands is crucial for understanding the electronic structure and correlated electronic states at ambient pressure, 
we apply a more accurate exchange-correlation functional to determine the interlayer binding and crystal splitting in LNO from the DFT perspective.

Our DFT calculations employ the Vienna ab-initio simulation package (VASP) code \cite{kresse1996efficient} with the projector augmented wave (PAW) method \cite{kresse1999ultrasoft}. The hybrid functionals that combine the Hartree-Fock (HF) and Kohn-Sham (KS) theories \cite{becke1993density} are used here. Typically, this can be more accurate compared to the semilocal GGA method \cite{exchange_correlation,Jacob_ladder}. Here, we implement the widely used HSE06 hybrid functional \cite{krukau2006influence}. We also tested other unscreened hybrid functionals such as PBE0 \cite{perdew1996rationale,ernzerhof1999assessment,adamo1999toward} and B3LYP \cite{stephens1994ab}. As discussed in the supplemental material (SM) \cite{SM}, all the hybrid functionals lead to similar results. The cutoff energy for expanding the wave functions into a plane-wave basis is set to be 500 eV. The energy convergence criterion is 10$^{−6}$
eV. All calculations are conducted using the primitive cell to save time. The $\Gamma$-centered 5$\times$5$\times$5 k-meshes are used. 
The band structures for the LP phase using the HSE06 functional is plotted in Figs. \ref{fig1}(d).
Compared to the GGA results (See Fig. S1 in SM \cite{SM}), the bonding $d_{z^2}$ bands are significantly pushed below $E_F$ by about $505$ meV ($\Gamma$ point in Fig. \ref{fig1}(d)).

\begin{figure}
	\begin{center}
		\fig{3.4in}{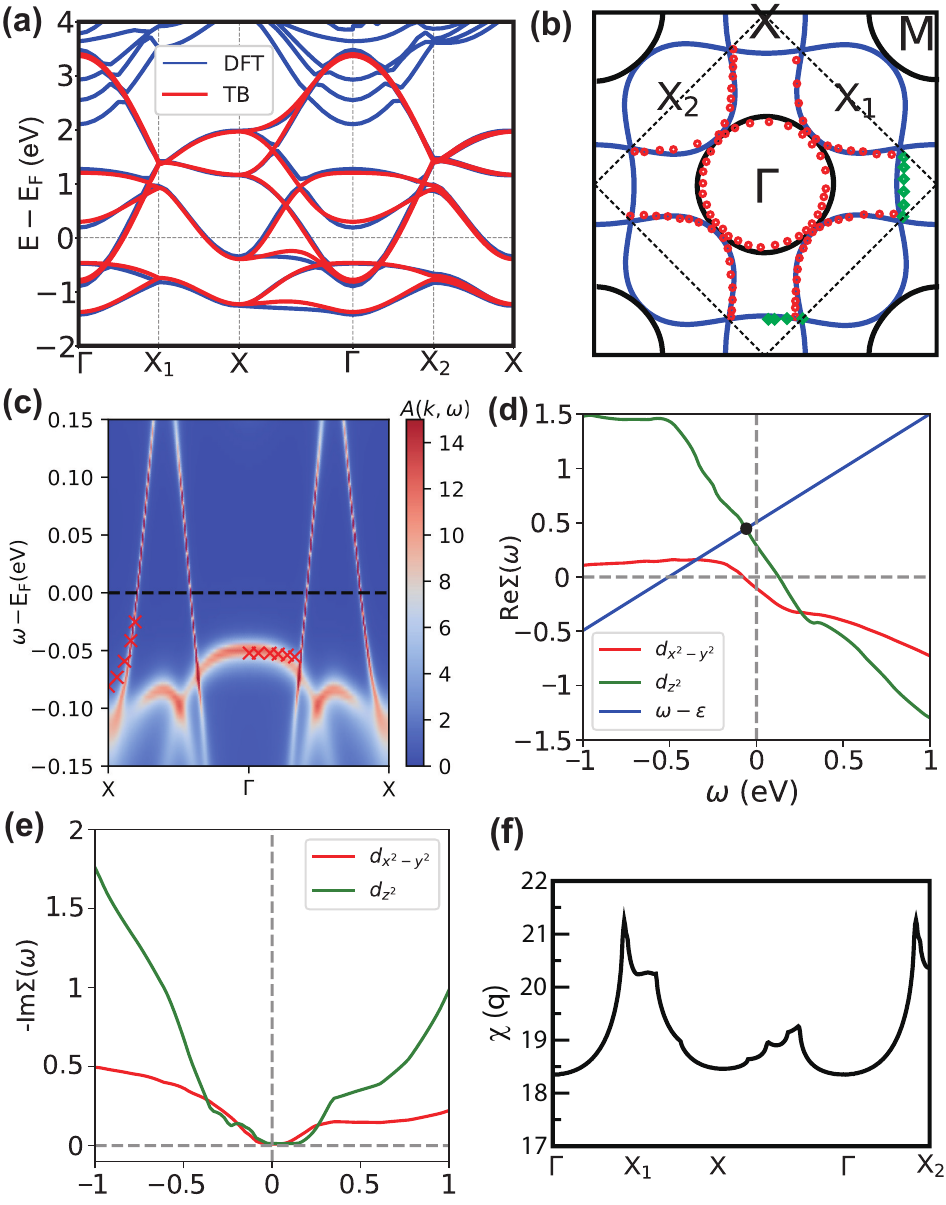}
		\caption{(a) The DFT band structure based on HSE06 hybrid functiona at ambient pressure (blue lines) and the corresponding TB bands (red lines). (b) The FSs at ambient pressure from the TB bands in (a). The red circles are experimental FSs from laser-based ARPES, while the green squares are data from synchrotron-based ARPES \cite{zhouxj}. Notice that the $\Gamma$ electron pocket nearly nests with the $X_1$ hole pocket at $Q=(\pi/2,\pi/2)$. (c) The spectral function $A(k,\omega)$ calculated using DMFT at $U$ = $5.5$ eV, $J_H$ = $0.65$ eV, and $\beta=50$ based on TB bands in (a). The red crosses in (c) are experimental data from ARPES measurements \cite{zhouxj}. (d) The real part of the electron self-energy from DMFT calculation. (e) The negative imaginary part of the electron self-energy from DMFT calculation. (f) The spin susceptibility $\chi(q)$ is plotted as a function of $q$, which peaks around $Q=(\pi/2, \pi/2)$ ($X_1$ or $X_2$).
			\label{fig2}}
	\end{center}
\end{figure}

Focusing on the partially occupied $e_g$ orbitals from the DFT calculations,  we use the Wannier90 code \cite{mostofi2008wannier90,marzari2012maximally} to fit the $e_g$ bands and extract the tight-binding (TB) model parameters. In the following discussion, we label the $d_{x^2-y^2}$ and $d_{z^2}$ orbital as $d_{x}$ and $d_{z}$ respectively. Due to the different octahedron environments, we define every $d_{x/z}$ orbital in its local octahedra coordinates rather than the global coordinates.
Because of the 4 atoms per unit cell, we obtain an 8-band TB model, whose Hamiltonian $H_{LP}$ and hopping parameters are given in the Tab. S1 in SM \cite{SM}. 
The dispersion of the TB $e_g$ bands is compared to the DFT calculations in Fig. \ref{fig2}(a), which confirms that the TB model provides a faithful description of the low-energy electron structure.

The Fermi surfaces (FSs) of the TB model in the LP phase are plotted in Fig. \ref{fig2}(b). There are four FSs, one electron pocket centered around $\Gamma$, another small electron pocket around $X$, and two hole pockets centering around $X_1$ and $X_2$, respectively. 
To directly compare to experimental findings,  we also plot the FS data obtained by angle-resolved photoemission spectroscopy (ARPES) recently in Fig. \ref{fig2}(b) \cite{zhouxj}. The red circles are data from laser-based ARPES and the green squares are data from synchrotron-based ARPES. Remarkably, our theoretical FSs obtained by DFT and HSE06 exchange-correlation functional show excellent agreement with the ARPES data.

\section{DMFT RESULTS}
To further the understanding of the electronic structure in the LP phase, we attempt a  theoretical description of the band dispersions measured by ARPES \cite{zhouxj}. To this end, it is necessary to include the electronic correlations beyond DFT, since the latter are known to produce the important renormalization of the band structure. We thus consider the multi-orbital Hubbard interactions,
\begin{align}
	\begin{split}
		H_I&= U\sum_{l,i,\eta} \hat{n}_{l,i,\eta \uparrow}\hat{n}_{l,i,\eta \downarrow}+(U-2J_H)\sum_{l,i,\eta\neq \eta^{\prime}}\hat{n}_{l,i,\eta}\hat{n}_{l,i,\eta^{\prime}}
		\\
		& -J_H\sum_{l,i,\eta\ne \eta^{\prime}} (\mathbf{S}_{l,i,\eta} \cdot  \mathbf{S}_{l,i,\eta'} 
		+d_{l,i,\eta\uparrow}^{\dagger}d_{l,i,\eta\downarrow}^{\dagger}d_{l,i,\eta^{\prime}\uparrow}d_{l,i,\eta^{\prime}\downarrow} )
	\end{split}
	\label{eq:HI}
\end{align}
where $l=t, b$ is the layer index, $i$ is the site index in each layer and $\eta=x,z$ is the orbital index. 
To obtain the correlation-induced band renormalization, we apply the dynamical mean field theory (DMFT) to the full Hamiltonian $H_{LP}+H_{I}$. Specifically, we implement the DMFT calculation using the open-source TRIQS \cite{triqs} package and its continuous-time quantum Monte Carlo as an impurity solver \cite{ctqmc}. The obtained DMFT results at $U=5.5$eV and $J_H=0.65$eV are plotted in Fig. \ref{fig2}(c) along the $X-\Gamma-X$ direction. The ARPES data \cite{zhouxj} are superimposed as red crosses in Fig. \ref{fig2}(c), which are very well described by the DMFT results. Despite the significant correlation-induced band renormalization necessary for the remarkable agreement with ARPES, we find that correlations do not cause appreciable inter-band carrier transfer and the interacting FSs are slightly changed from Fig. \ref{fig2}(b). 

For a more quantitative analysis, we calculated the real (Re$\Sigma(\omega)$) and negative imaginary (-Im$\Sigma(\omega)$) parts of the electronic self-energy, as shown in Figs. \ref{fig2}(d,e). The Re$\Sigma(\omega)$  has already been subtracted by the double-counting term $\Sigma_{dc}$ and the chemical potential $\mu$. With the help of the Re$\Sigma(\omega)$, we can easily observe how the band top of the $d_{z}$ bonding state at the $\Gamma$ point changes. For a single-particle state with energy $\epsilon$ in DFT, the energy $\omega$ in DMFT for this state is determined by the following equation:
\begin{align}
    \omega-\epsilon=\rm{Re}\Sigma(\omega).
\end{align}
Therefore, we can determine the energy level position in DMFT by locating the intersection of the straight line 
$\omega-\epsilon$ and the Re$\Sigma(\omega)$. According to the DFT results, $\epsilon=-0.505$eV, and the main orbital component of this energy level is $d_z$. We mark the intersection point with a black dot, as shown in Fig. \ref{fig2}(d). The $\omega$ value corresponding to this point ($\sim-0.05$eV) quantitatively matches the position of the band top in Fig. \ref{fig2}(c). From the Im$\Sigma(\omega)$, we can see that as $\omega$ approaches 0, the Im$\Sigma(\omega)$ also approaches 0. This indicates that at this temperature ($\beta=50$, approximately 230K), the system is in a Fermi liquid phase \cite{PhysRevLett.101.166405}, which is different from the DMFT results from GGA functional, as detailed in the Ref. \cite{PhysRevB.109.115114} and SM \cite{SM}.


These overall agreements with the experimental findings reassures that our new DFT electronic structure captured by the TB models plus the multi-orbital Hubbard interactions provide a reliable description for the physics of La$_3$Ni$_2$O$_7$ in its LP phase.

However, it is also important to emphasize that the energy bands calculated using the traditional GGA functional do not match well with the ARPES results even after considering correlation effects. 
Previously, many attempts using the DFT+U calculations based on the GGA functional have applied to the LP phase \cite{zhouxj}. Although DFT+U can push the bands below the Fermi level, it fails to capture the band renormalization effect as shown in Fig. S3(b) in SM \cite{SM}. We also apply the DMFT calculation with the TB inputs from traditional GGA functional. Although this method can capture the band renormalization effect, it fails to push the bands below the Fermi level as shown in Fig. S3(a). The more detailed information can also be seen in SM \cite{SM}.

\section{MAGNETIC STRUCTURE}
Surprisingly, we find that the $\Gamma$ electron FS nearly nests with the $X_1$ hole FS, as marked in Fig. \ref{fig2}(b). The nesting between electron and hole FS pockets implies divergent electronic susceptibility at the nesting wave vector and potential symmetry-breaking density wave order at low temperatures. The calculated
spin susceptibility $\chi(q)$ using TB bands from Fig. \ref{fig2}(a) is shown in Fig. \ref{fig2}(f) at different wave vectors, which indeed peaks around the near-nesting wave vector
$Q=(\pi/2, \pi/2)$. This intriguing spin susceptibility feature is consistent with experimental findings \cite{rixs,musr,musr2,nmr}.

\begin{figure}
	\begin{center}
		\fig{3.4in}{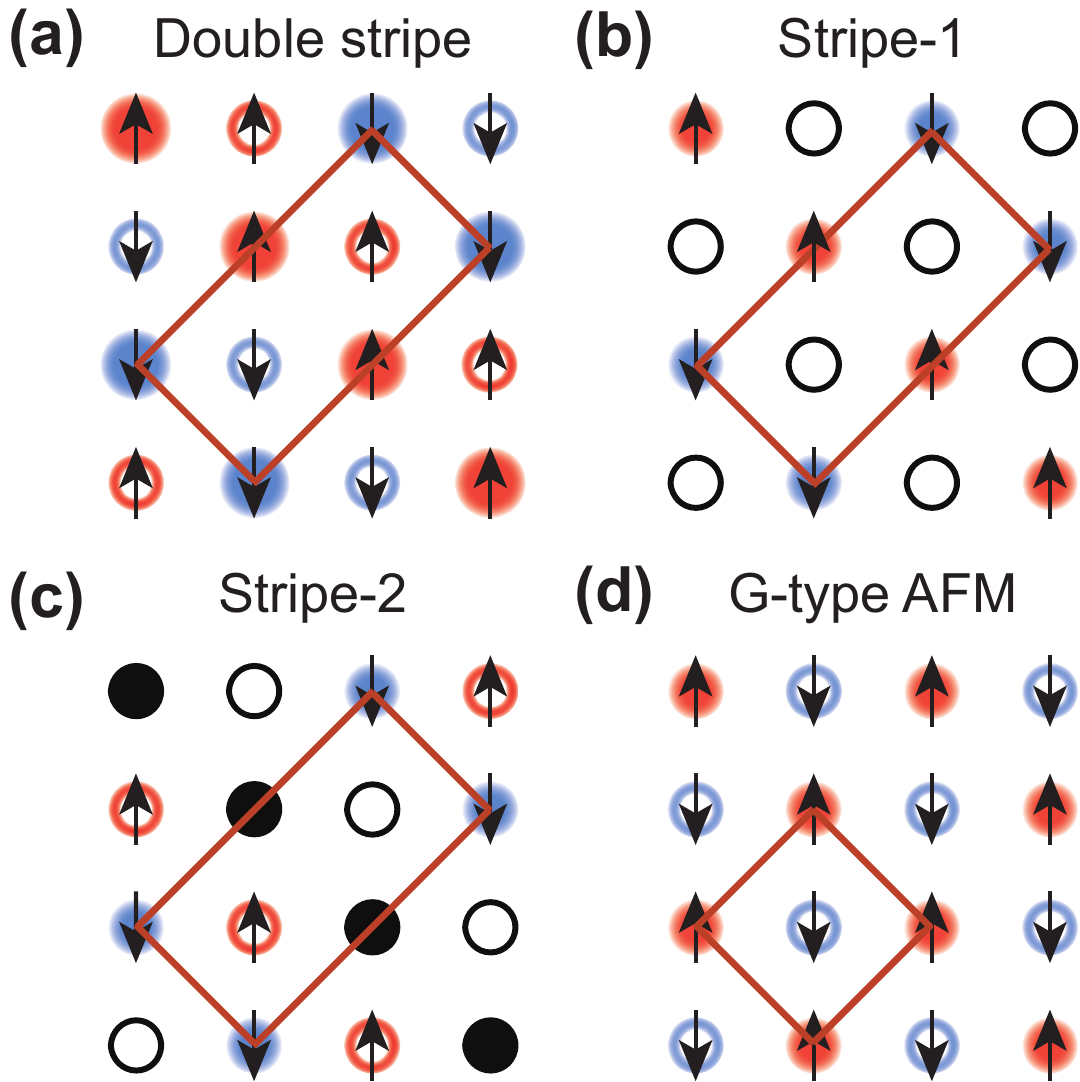}
		\caption{(a)-(c) The schematic illustration for the three considered ($\pi/2$, $\pi/2$) stripe spin order in LNO lattice. Specifically, (a) The double-stripe order, (b) The stripe-1 order and (c) The stripe-2 order. Since LNO has two different types of sublattice, we label the open and solid circles representing A and B sites respectively. From the DFT (HSE06)+U calculation, a charge-spin intertwined order emerges in (a). The alternating sizes of the dots in (a) represent the charge order. (d) The schematic illustration for G-type AFM ordering at ($\pi$, $\pi$). The red, blue and black dots represent spin up, spin down and non-magnetic Ni atoms, respectively. The brown rectangle represents the magnetic unit cell.}
			\label{fig3}
	\end{center}
\end{figure}

In order to further confirm this, we apply the mean-field calculation to the above multi-orbital Hubbard interaction. To simplify the calculation, we focus on the spin channel (the charge channel does not develop at this mean-field level):
\begin{align}
    \begin{split}
        H_{MF}=H_{LP}-&\sum_{l,i,\eta}[\frac{U}{2} m_{l,i,\eta}+\frac{J_H}{2}\sum_{\zeta\neq\eta}m_{l,i,\zeta}]\hat{m}_{l,i,\eta} \\
        +&\sum_{l,i,\eta} [\frac{U}{4} m_{l,i,\eta}+\frac{J_H}{4}\sum_{\zeta\neq\eta}m_{l,i,\zeta}]m_{l,i,\eta},
    \end{split}
\end{align}
where $\hat{m}_{l,i,\eta}=\hat{d}_{l,i,\eta\uparrow}^{\dagger}\hat{d}_{l,i,\eta\uparrow}-\hat{d}_{l,i,\eta\downarrow}^{\dagger}\hat{d}_{l,i,\eta\downarrow}$ is the magnetic moment operator of orbital $\eta$ at layer $l$, site $i$. 
For the mean-field ansatzes, we take three spin-density waves (SDWs) ordering at $Q$ into account, the double stripe (Fig. \ref{fig3}(a)), stripe-1(Fig. \ref{fig3}(b)), stripe-2 (Fig. \ref{fig3}(c)). Additionally, noticing that the bilayer Ni is strongly coupled, we always assume the  $m_{t,i,\eta}=-m_{b,i,\eta}$. We also study the G-type AFM (Fig. \ref{fig3}(d)) ordering at $(\pi,\pi)$ to complete our analysis.
After the self-consistent calculations as a function of $U$ with $J_H=0.1U$, we find three phases are highly competing except for the unstable stripe-2 phase. 
The mean-field energies $\langle H_{MF} \rangle$ of the three orders are listed in Tab. \ref{tab;MF magnetic energy}. 
When $U$ is small, G-type AFM is the lowest energy state. However, when $U>2.4$eV, the double stripe becomes dominant. Hence, this simple Hatree-Fock calculation confirms the double stripe order is the magnetic ground state at a large $U$ limit.


\begin{table}[!htbp] 
\centering
\caption{Total energy per Ni atom of the different magnetic orders from the mean-field calculation.}
\begin{ruledtabular}
\begin{tabular}{cccccc}
Energy (eV)&$U=2$&$U=2.4$&$U=2.7$&$U=3$&$U=3.4$ \\
Double stripe&-1.061&-1.082&-1.118&-1.168&-1.250 \\
Stripe-1&-1.061&-1.073&-1.094&-1.126&-1.184 \\
G-type AFM&-1.063&-1.086&-1.117&-1.157&-1.223
\end{tabular}
\end{ruledtabular}
\label{tab;MF magnetic energy}
\end{table}

On the other hand, the above mean field cannot fully take care of the charge channel, which needs to be feedback to the electronic structure. A better way is using the simplified rotation invariant approach to the DFT+U, introduced by Dudarev $et$ $al.$ \cite{dudarev1998electron}. Our calculations here are still based on the HSE06 hybrid functional. We also consider the above three stripe states and G-type.
We find that within the range of $U=0$eV to $U=4$eV, only the double-stripe configuration remains stable. The other two configurations spontaneously converge to the double stripe order. Therefore, we can be convinced that the double-stripe configuration is the magnetic ground state. 
Interestingly, the double-stripe order found here is a spin-charge intertwined state. 
The Ni atoms in the double-stripe order are divided into two groups with different magnetic moments and different charge densities. The larger charge atoms host relatively larger magnetic moments, as illustrated in Fig. \ref{fig3}(a). Therefore, a charge density wave order at $(\pi, \pi)$ emerges from the spin density wave ordering at $Q=(\pi/2, \pi/2)$.  


\begin{table}[!htbp] 
\centering
\caption{Total energy per Ni atom of the different magnetic orders in DFT (HSE06)+U calculation.}
\begin{ruledtabular}
\begin{tabular}{cccccc}
Energy (eV)&$U=0$&$U=1$&$U=2$&$U=3$&$U=4$ \\
Double stripe&-54.750&-54.429&-54.135&-53.869&-53.633 \\
G-type AFM&-54.518&-54.171&-53.856&-53.577&-53.340
\end{tabular}
\end{ruledtabular}
\label{tab;magnetic energy}
\end{table}


\section{SUMMARY}
Although the main focus of our paper is the electronic structure and magnetism of the LP phase, we would like to briefly mention the understanding of superconductivity in the HP using the HSE06 functional. In previous studies \cite{yaodx,dagotto1,wangqh,Kuroki,guyh,zhanggm,werner,yangf,wucj,dagotto2,yangyf2,ryee2023critical,kun_cpl,PhysRevB.108.L201121}, it has been suggested that the bonding state of the $d_z$ orbital plays a crucial role in superconducting pairing, as it remains active at the Fermi surface. There is also a debate regarding the $s^{\pm}$-wave \cite{wangqh,guyh,dagotto2} and $d$-wave pairing symmetries \cite{kun_cpl,PhysRevB.108.L201121}. However, our use of the HSE06 functional indicates that in the LP phase, the bonding state of $d_z$ is fully occupied, meaning it no longer appears at the Fermi surface. In fact, we also calculated the band structure of the HP phase using the same method (with the crystal structure obtained from \cite{meng_wang}), as shown in Fig. S5 in SM \cite{SM}. Due to band unfolding, the band top of the bonding state of $d_z$
is located at the M point, and it is clearly still below the Fermi level. Therefore, we believe that based on our band structure, the superconducting order parameter is likely to exhibit $d$-wave symmetry, which is consistent with our previous theoretical analysis \cite{kun_cpl}.  Currently, there are first-principles techniques that, by combining DFT-calculated band structures with pairing assumptions, provide further insight \cite{csire2018nonunitary}. It is also worth further investigating the impact of band structure on the superconducting ground state using this approach.

In summary, we presented a systematic study on the correlated electronic structure and magnetic properties in LP bilayer La$_3$Ni$_2$O$_7$. Utilizing advanced exchange-correlation functionals, 
new electronic structures of LNO have been identified from the DFT calculations for the LP phases. In sharp contrast to previous GGA functional results, the bilayer coupling and the $e_g$ crystal fields are greatly enhanced. This enhancement pushes the bonding $d_{z^2}$ band down to lie below the Fermi level. The obtained Fermi surface structure in the LP phase matches well the recent ARPES measurements performed at ambient pressure. 
Beyond the DFT calculation, we apply the DMFT
to study the effects of multi-orbital Hubbard interactions using the DFT 
(HSE06) derived TB models in the low-pressure phase. The renormalized band structures quantitatively match the bands detected by the ARPES-measured spectral functions. 
Moreover, using the new TB model, the calculated spin susceptibility exhibits near nesting peak structures around the wave vector $Q=(\pi/2,\pi/2)$, which is consistent with recent RIXS resolved magnon dispersion \cite{rixs} and additional evidence from the $\mu$SR and NMR experiments \cite{musr,musr2,nmr}. Furthermore, using the DFT (HSE06)+U method and mean-field decoupling, we confirm that the ground state of the magnetic structure should be the double stripe order. A charge density wave intertwined with this SDW is also found, which calls for experimental verification.
Therefore, we think our new TB model plus multi-orbital Hubbard interactions can provide a more faithful description of the electronic and magnetic structure in LNO.


\textbf{Acknowledgement} We acknowledge the support by the Ministry of Science and Technology  (Grant No. 2022YFA1403900), the National Natural Science Foundation of China (Grant No. NSFC-11888101, No. NSFC-12174428, No. NSFC-11920101005), the Strategic Priority Research Program of the Chinese Academy of Sciences (Grant No. XDB28000000, XDB33000000), the New Cornerstone Investigator Program, and the Chinese Academy of Sciences Project for Young Scientists in Basic Research (2022YSBR-048). Z.W. is supported by the U.S. Department of Energy, Basic Energy Sciences Grant No. DE-FG02-99ER45747.



\bibliography{main.bbl}

\clearpage
\onecolumngrid
\begin{center}
\textbf{\large Supplemental Material: The electronic and magnetic structures of bilayer La$_3$Ni$_2$O$_7$ at ambient pressure}
\end{center}

\setcounter{equation}{0}
\setcounter{figure}{0}
\setcounter{table}{0}
\setcounter{page}{1}
\makeatletter
\renewcommand{\theequation}{S\arabic{equation}}
\renewcommand{\thefigure}{S\arabic{figure}}
\renewcommand{\thetable}{S\arabic{table}}

\twocolumngrid

\subsection{Exchange-correlation functional, GGA, and Hybrid functional}
\begin{figure}[htbp!]
	\begin{center}
		\fig{3.4in}{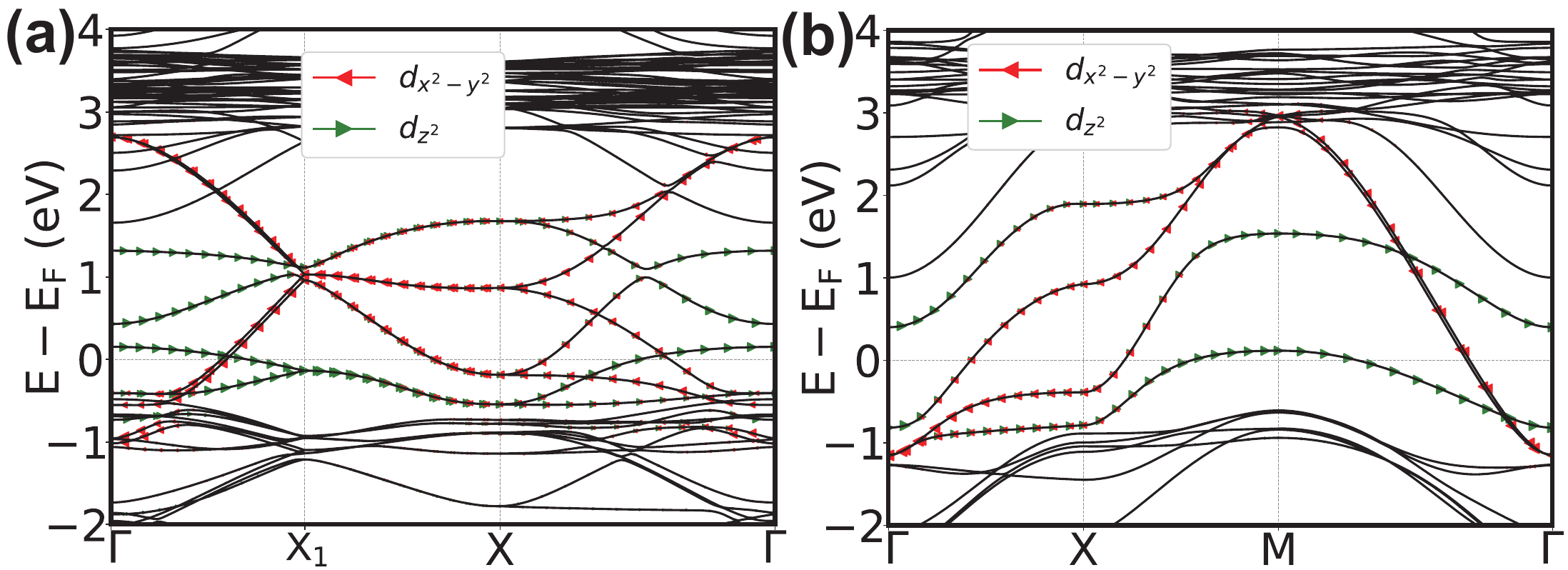}
		\caption{LNO electronic structure obtained using GGA exchange-correlation functional. (a) GGA functional calculated band dispersions in the LP phase. (b) GGA functional calculated band dispersions in the HP phase.
			\label{pbe}}
	\end{center}
\end{figure}

DFT is a powerful computational method to investigate the electronic structure of solids. 
The basic computational methods behind DFT calculations are the Kohn-Sham equations, which provide a systematic way of mapping the many-body system onto a single-body problem \cite{martin2020electronic}.
More precisely, the total energy of the system in Kohn-Sham theory is written as a functional of electron density $\rho(\mathbf{r})$,
\begin{eqnarray}
    E[\rho]=T_s[\rho]+\int d\mathbf{r} v_{ext}(\mathbf{r})\rho(\mathbf{r})+E_H[\rho]+E_{xc}[\rho]
\end{eqnarray}
The $T_s[\rho]$ is the Kohn-Sham kinetic energy, $v_{ext}(\mathbf{r})$ is the external potential like the electron-nuclei interaction, and $E_H[\rho]$ is the Hatree energy. The $E_{xc}[\rho]$ is the exchange-correlation functional. $T_s$, $v_{ext}$, $E_H[\rho]$ can be straightforwardly calculated. The only unknown in Kohn-Sham theory is the exchange-correlation functional $E_{xc}$ \cite{martin2020electronic,Jacob_ladder}.
Developing a precise $E_{xc}$ is one central goal in the DFT framework. Hence, implementing a precise $E_{xc}$ is a crucial step for band structure calculation.
The most commonly used $E_{xc}$ in solid-state physics is the GGA or its PBE version \cite{perdew1996generalized}. 
The GGA band structures of LNO in the LP and HP phases are plotted in Fig. \ref{pbe}. In both cases, the bonding $d_z$ bands cross the Fermi level, which is due to the underestimated exchange correlations. 

\begin{figure}[htbp!]
	\begin{center}
		\fig{3.4in}{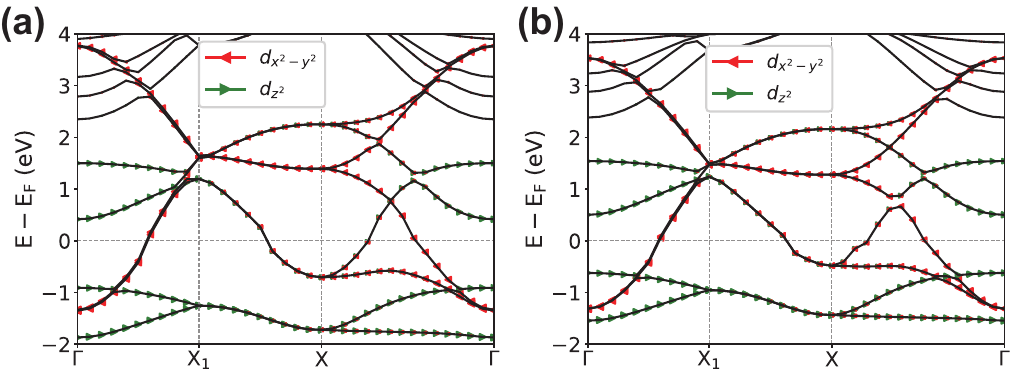}
		\caption{LNO band structure calculated using the hybrid functional. (a) PBE0 functional calculated band dispersions in the LP phase. (b) B3LYP functional calculated band dispersions in the LP phase.
			\label{hybrid}}
	\end{center}
\end{figure}

Going beyond GGA, a more advanced exchange-correlation functional $E_{xc}$ is the hybrid functional. The exchange part or correlation part of $E_{xc}$ is a linear combination of the Hatree-Fock (HF) exchange and semilocal potential such as GGA. The hybrid functionals can be divided into two families according to the interelectronic range at which the HF exchange is applied: screened (range-separated) and unscreened. The screened hybrid functional like sophisticated HSE06 which mixes GGA and HF in the short-range exchange energy is often used in solid state calculations. It provides great accuracy for band gaps \cite{exchange_correlation}.
There are also other unscreened hybrid functionals like the PBE0 \cite{perdew1996rationale,ernzerhof1999assessment,adamo1999toward} and B3LYP \cite{stephens1994ab}, which have slower convergence than HSE06 and also been tested. PBE0 is similar to HSE06 but mixes both the short-range and the long-range contributions. B3LYP mixes LDA, HF, and Becke88 in the exchange energy. The correlation energy is also a mixture of LYP and VWN3. The PBE0 and B3LYP band structures are shown in Fig. \ref{hybrid}, which are similar to HSE06 in the main text. 

\subsection{DMFT and DFT+U}

Although the hybrid functional provides an accurate description of band structure, the local Hubbard interactions beyond DFT should be treated using other many-body techniques. One commonly used method is DMFT. In the main text, we presented a DMFT study of the spectral function in the LP phase. The basic assumption of DMFT is that the self-energy only depends on the frequency $\Sigma(\omega)$. We use the open-source TRIQS DMFT package \cite{triqs} and its CTQMC solver \cite{ctqmc}. In the LP phase, there are four Ni atoms. To simplify our discussion, we treat them equally in the DMFT calculation. During each DMFT self-consistent calculation, a $40\times40$ Brillouin zone mesh is used for the lattice Green's function calculation. 

\begin{figure}[htbp!]
	\begin{center}
		\fig{3.4in}{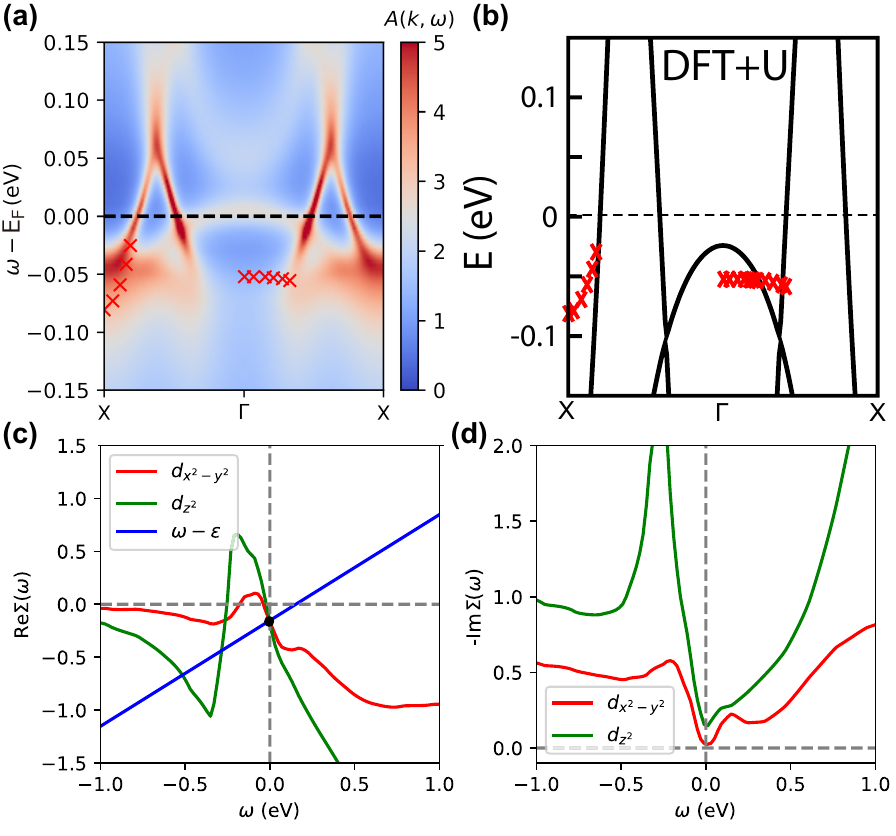}
		\caption{(a) The spectral function $A(k,\omega)$ calculated using DMFT at $U$ = $5.5$ eV, $J_H$ = $0.65$ eV, and $\beta=50$ based on TB bands fitted from DFT bands of ordinary GGA functional. The parameters used here are same as the main text. (b) The DFT+U band dispersion using GGA functional at $U=4$ eV. The calculated band dispersion cannot describe the experiment data (red crosses) owing to the band position being incorrect in (a) and missing additional renormalization effects in (b). (c) The real part of the electron self-energy from DMFT calculation. (d) The negative imaginary part of the electron self-energy from DMFT calculation.
			\label{figldau}}
	\end{center}
\end{figure}

Therefore, an immediate question is whether using only the traditional GGA functional, after considering many-body interactions, can yield band structures consistent with experiments. Hence, we also performed DMFT calculations using a TB model fitted from previous band calculations \cite{yaodx} with the GGA functional, as shown in Fig. \ref{figldau}(a). It is evident that even with interactions included, the $d_{z}$ bonding state is not fully pushed below the Fermi level. From Fig. \ref{figldau}(c), by examining the real part of the self-energy and using the method mentioned in the main text, we can also determine that the band top of the $d_z$ bonding state ($\epsilon=0.153$eV from GGA-based DFT calculation) is exactly at the Fermi level, which is consistent with the results of the spectral function. From Fig. \ref{figldau}(d), we can see that at $\omega=0$, the imaginary part of the self-energy for both orbitals is nonzero at $\beta=50$ ($\sim$ 230K), indicating that the DMFT result from GGA functional corresponds to a frozen moment phase \cite{PhysRevLett.101.166405} at this temperature, consistent with previous calculations \cite{PhysRevB.109.115114}. However, this differs from the result obtained from the HSE06 functional in the main text.

The DFT+U is another convenient approach \cite{martin2020electronic}. However, DFT+U only shifts the crystal field splitting without renormalizing the hopping parameters or the bandwidths. In previous works, it was proposed that DFT+U (using GGA functional) can capture the ARPES measurements of the band dispersion in ambient pressure \cite{zhouxj}. We also calculate the band structure of the LP phase using the simplified rotation invariant approach to the DFT+U introduced by Dudarev $et$ $al.$ \cite{dudarev1998electron} based on GGA functional. As shown in Fig. \ref{figldau}(b), the DFT+U results with $U$=4eV show significant deviation from the ARPES measurements, particularly in terms of the renormalization effect, despite the band positions being nearly correct.

\subsection{Hamiltonian and Tight-binding parameters}
The TB model for LNO in the LP phase can be derived in the basis (the spin index is omitted here) $(d_{tA\textbf{k}}^{x},d_{tA\textbf{k}}^{z},d_{tB\textbf{k}}^{x},d_{tB\textbf{k}}^{z},d_{bA\textbf{k}}^{x},d_{bA\textbf{k}}^{z}, d_{bB\textbf{k}}^{x},d_{bB\textbf{k}}^{z})$ as 

\begin{align}
H_{LP} & ({\textbf{k}})=\left(\begin{array}{cc}
H_{t}({\textbf{k}}) & H_{\perp}({\textbf{k}})\\
H_{\perp}^{\dagger}({\textbf{k}}) & H_{b}({\textbf{k}})
\end{array}\right),\label{eq:tb-lp}
\end{align}
where $H_{b}({\textbf{k}})=H_{t}({\textbf{k}})$ and $H_{\perp}({\textbf{k}})=H_{\perp}^{\dagger}({\textbf{k}})$. They take the structures of

\begin{align}
H_{t}({\textbf{k}})=\left(\begin{array}{cccc}
H_{11}(\textbf{k}) & 0 & H_{13}(\textbf{k}) & H_{14}(\textbf{k})\\
0 & H_{22}(\textbf{k}) & H_{14}(\textbf{k}) & H_{24}(\textbf{k}) \\
H^{*}_{13}(\textbf{k}) & H^{*}_{14}(\textbf{k}) & H_{11}(\textbf{k}) & 0 \\
H^{*}_{14}(\textbf{k}) & H^{*}_{24}(\textbf{k}) & 0 & H_{22}(\textbf{k}) 
\end{array}\right), 
\end{align}

\begin{align}
H_{\bot}({\textbf{k}})=\left(\begin{array}{cccc}
t_{\bot}^{x} & 0 & 0 &  H_{18}(\textbf{k})\\
0 &  t_{\bot}^{z} &  H_{18}(\textbf{k}) & 0 \\
0 &  H^{*}_{18}(\textbf{k}) &  t_{\bot}^{x} & 0 \\
H^{*}_{18}(\textbf{k}) & 0 & 0 & t_{\bot}^{z} 
\end{array}\right),
\end{align}
with $H_{11/22}(\textbf{k})=\epsilon^{x/z}+2({t'_{2}}^{x/z}\cos k_{X}+{t_{2}}^{x/z}\cos k_{Y})+4t_{5}^{x/z}\cos k_{X}\cos k_{Y}$, $H_{13/24}(\textbf{k})=2\cos \frac{1}{2}k_{X}({t_{1}}^{x/z}e^{i\frac{1}{2}k_{Y}}+{t'_{1}}^{x/z}e^{-i\frac{1}{2}k_{Y}})$ and $H_{14/18}(\textbf{k})=2i\sin \frac{1}{2}k_{X}({t_{3/4}}^{xz}e^{i\frac{1}{2}k_{Y}}-{t'_{3/4}}^{xz}e^{-i\frac{1}{2}k_{Y}})$. Here, $k_X=k_x+k_y$ and $k_Y=-k_x+k_y$ are vectors defined in the enlarged unit cell. The hopping parameters in the TB model for the bilayer eight-orbital model in the LP phase are listed in Tab. \ref{tab:hop2}. The corresponding schematic illustration is shown in Fig. \ref{figTB}. We add the third nearest neighbor hopping parameters $t_{5}^{x/z}$ in the diagonal terms in order to achieve a better fitting.

\begin{figure}[htbp!]
	\begin{center}
		\fig{3.4in}{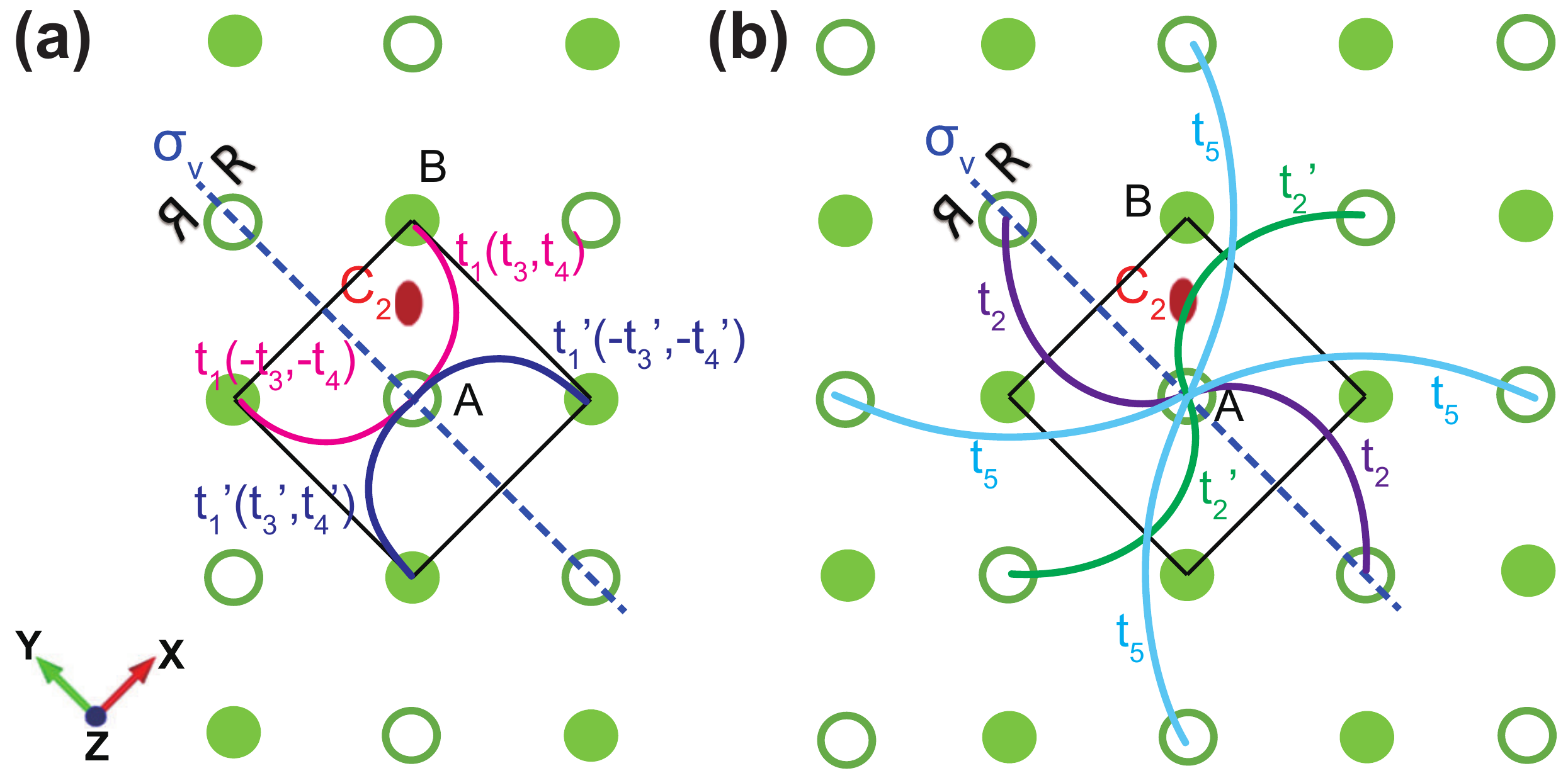}
		\caption{(a), (b) The hopping parameters used in TB of LP LNO.
			\label{figTB}}
	\end{center}
\end{figure}

\begin{table}[htbp!]
\begin{tabular}{ccccccccc}
\hline \hline 
$t_{1}^{x}$ & ${t'_{1}}^{x}$ & $t_{1}^{z}$ & ${t'_{1}}^{z}$ & $t_{2}^{x}$ & ${t'_{2}}^{x}$ & $t_{2}^{z}$ & ${t'_{2}}^{z}$ & $t_{3}^{xz}$
\tabularnewline\hline 
-0.519 & -0.525 & -0.123 & -0.104 & 0.069 & 0.08 & -0.019 & -0.018 & 0.271 \tabularnewline \hline 
 ${t'_{3}}^{xz}$ & $t_{4}^{xz}$ & ${t'_{4}}^{xz}$ & $t_{5}^{x}$ & $t_{5}^{z}$ & $t_{\bot}^{x}$ & $t_{\bot}^{z}$ & $\epsilon^{x}$ & $\epsilon^{z}$  
\tabularnewline \hline 
0.258 & -0.033 & -0.026 & -0.051 & -0.013 & 0.017 & -0.835 & 1.196 & 0.039 
\tabularnewline
\hline \hline
\end{tabular}
\caption{\label{tab:hop2} Hopping parameters in the TB Hamiltonian (unit here is eV) in the LP phase.  $\epsilon^{x},\epsilon^z$ are site energies for Ni $d_{x^2-y^2}$ and $d_{3z^2-r^2}$ orbitals, respectively.}
\end{table}

\subsection{HSE06 band structure of HP phase}
We also calculated the band structure and Fermi surface of the HP phase using the HSE06 method, as shown in Fig. S5. As can be seen, due to the doubling of the first Brillouin zone, the top of the $d_{z}$ bonding state, which is located at the $\Gamma$ point in the LP phase, is unfolded back to the M point in the HP phase. However, the most important point is that $d_{z}$ bonding state remains fully occupied.

\begin{figure}[htbp!]
	\begin{center}
		\fig{3.4in}{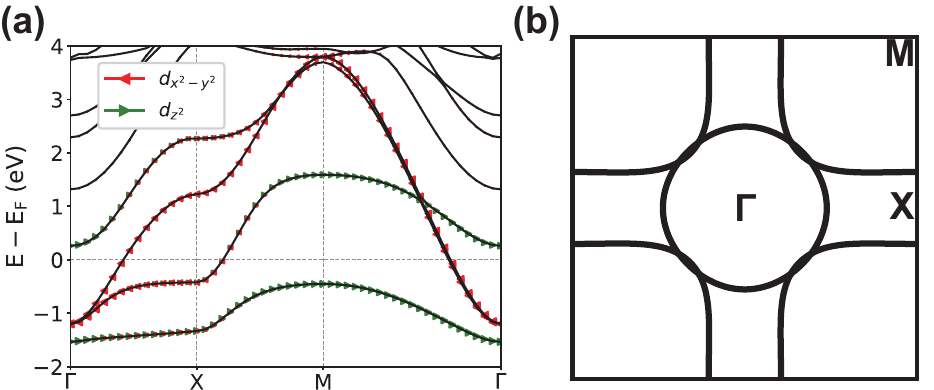}
		\caption{(a) The HSE06 band structure of HP phase. (b) The HSE06 Fermi surface of HP phase.
			\label{figTB}}
	\end{center}
\end{figure}

\end{document}